\documentclass[conference]{IEEEtran}
\usepackage{cite}
\usepackage{amsmath,amssymb,amsfonts}
\usepackage{algorithmic}
\usepackage{graphicx}
\usepackage{textcomp}
\usepackage{xcolor}
\usepackage{booktabs}
\usepackage{siunitx}
\usepackage{hyperref}
\usepackage{float}
\usepackage{multirow}

\def\BibTeX{{\rm B\kern-.05em{\sc i\kern-.025em b}\kern-.08em
    T\kern-.1667em\lower.7ex\hbox{E}\kern-.125emX}}

\begin{document}

\title{Torsion-Space Diffusion for Protein Backbone Generation with Geometric Refinement}

\author{\IEEEauthorblockN{1\textsuperscript{st} Adwait Shelke}
\IEEEauthorblockA{\textit{Dept. of Mechanical Engineering} \\
\textit{Indian Institute of Technology Bombay}\\
Mumbai, India \\
23B2119@iitb.ac.in}
\and
\IEEEauthorblockN{2\textsuperscript{nd} Lakshaditya Singh}
\IEEEauthorblockA{\textit{Dept. of Mechanical Engineering} \\
\textit{Indian Institute of Technology Bombay}\\
Mumbai, India \\
23B2230@iitb.ac.in}
\and
\IEEEauthorblockN{3\textsuperscript{nd} Divyansh Agrawal}
\IEEEauthorblockA{\textit{Dept. of Mechanical Engineering} \\
\textit{Indian Institute of Technology Bombay}\\
Mumbai, India \\
23B2204@iitb.ac.in}
}

\maketitle

\begin{abstract}
Building new protein structures is a major hurdle in computational biology, impacting everything from drug design to creating new enzymes. The problem with standard diffusion models is that they usually work in Cartesian coordinates. This often breaks the rules of geometry—specifically, it messes up bond lengths and angles. We took a different approach: a Torsion-Space Diffusion Model. Instead of guessing 3D coordinates directly, our model predicts torsion angles ($\phi, \psi, \omega$). By design, this guarantees that the local geometry stays perfect (keeping bond lengths at exactly 3.8\AA). To handle the overall shape, we added a refinement step that tweaks the Radius of Gyration (Rg) without breaking those bond constraints. Testing this on a standard protein dataset, we hit 100\% bond length accuracy (mean 3.800\AA, $\sigma \approx 0.0002$\AA) and saw a massive jump in global compactness (cutting Rg error down to 18.6\%) compared to Cartesian baselines. Basically, by combining a Transformer-based denoiser with a geometric converter, we created a solid framework for generating physically valid protein backbones.
\end{abstract}

\begin{IEEEkeywords}
Protein Design, Diffusion Models, Torsion Angles, Geometric Deep Learning, Generative AI
\end{IEEEkeywords}

\section{Introduction}
Proteins run biological systems, and their function depends entirely on their 3D shape. If we can design new protein structures from scratch (\textit{de novo}), we open up huge possibilities for medicine and synthetic biology. Recently, Deep generative models, especially Denoising Diffusion Probabilistic Models (DDPMs) \cite{ho2020denoising}, have become the go-to tools for this \cite{trippe2022diffusion, watson2023rfdiffusion}.

But here is the catch: modeling proteins as a cloud of points in Cartesian space is messy. Proteins aren't just random points; they are kinematic chains with strict rules for bond lengths and angles. When Cartesian diffusion models try to generate these, they often fail to respect those hard constraints. You end up with "broken" structures where the chemistry doesn't make sense. For example, adding standard Gaussian noise to Cartesian coordinates essentially destroys the bond topology that keeps a protein stable.

To fix this, we built a diffusion framework that works in the protein's natural coordinate system: torsion angles. We diffuse and denoise the dihedral angles ($\phi, \psi, \omega$) and then rebuild the 3D coordinates using a differentiable forward kinematics layer. This ensures every structure we generate is locally valid. We also tackled a common issue with angle-based models—poor global compactness—by adding a constrained refinement step that optimizes the Radius of Gyration (Rg).

In short, our contributions are:
\begin{itemize}
    \item A diffusion architecture based on Transformers that works purely in torsion space.
    \item A differentiable converter that turns torsion angles into coordinates while locking bond lengths at 3.8\AA.
    \item An iterative refinement algorithm that fixes global compactness (Rg) without breaking local bonds.
    \item Experimental proof that our method produces geometrically superior structures compared to Cartesian baselines.
\end{itemize}

\section{Related Work}
\subsection{Generative Models for Proteins}
In the early days, protein generation was mostly about assembling fragments and minimizing energy (like Rosetta). Later, deep learning methods like Variational Autoencoders (VAEs) and GANs entered the picture. They worked to an extent, but often suffered from mode collapse or missed the complex, long-range dependencies that define protein structures.

\subsection{Diffusion Models}
Diffusion models \cite{ho2020denoising} have changed the game in image generation and are now doing the same for molecular structures. RFdiffusion \cite{watson2023rfdiffusion} is a standout example, fine-tuning RoseTTAFold for structure denoising. While powerful, many of these models work in $SE(3)$ space or Cartesian coordinates, which means they need extra loss functions just to enforce basic bond constraints. Our work aligns more with approaches like \cite{trippe2022diffusion}, but we simplify things by focusing strictly on torsion space with a lighter Transformer backbone.

\section{Methodology}

\subsection{Data Preparation}
We used protein structures from the Protein Data Bank (PDB). Our prep pipeline was straightforward:
\begin{enumerate}
    \item \textbf{Extraction}: We pulled the backbone coordinates ($N, C_\alpha, C$).
    \item \textbf{Filtering}: We cut proteins longer than 128 residues to keep batching consistent.
    \item \textbf{Torsion Conversion}: We converted Cartesian coordinates into torsion angles ($\phi, \psi, \omega$) using standard geometric formulas.
    \item \textbf{Normalization}: We normalized the angles to fit the range $[-\pi, \pi]$.
\end{enumerate}
This gave us a dataset including standard benchmarks like Ubiquitin (1ubq), Crambin (1crn), Adenylate Kinase (1ake), and Lysozyme (2lyz).

\subsection{Torsion-Space Diffusion Model}
At the core of our method is a diffusion model trained to reverse a noise process applied to torsion angles.

\subsubsection{Forward Process}
We define a forward process that slowly adds Gaussian noise to the torsion angles $x_0$ over $T$ timesteps, creating a sequence $x_1, \dots, x_T$. We used a cosine schedule for $\beta_t$ to keep the signal intact for longer.
\begin{equation}
    q(x_t | x_{t-1}) = \mathcal{N}(x_t; \sqrt{1-\beta_t}x_{t-1}, \beta_t \mathbf{I})
\end{equation}
Here, $x_t \in \mathbb{R}^{L \times 3}$ represents the torsion angles at timestep $t$.

\subsubsection{Denoising Network Architecture}
For the denoising network, we chose a Transformer architecture to handle the long-range dependencies along the protein chain.
\begin{itemize}
    \item \textbf{Input}: Noisy angles $x_t \in \mathbb{R}^{L \times 3}$ and the timestep embedding $t$.
    \item \textbf{Embedding}: We added sinusoidal positional embeddings to the inputs.
    \item \textbf{Transformer Blocks}: 4 layers of self-attention blocks:
    \begin{itemize}
        \item Hidden dimension: 256
        \item Attention heads: 4
        \item Feed-forward dimension: 1024
        \item Dropout: 0.1
    \end{itemize}
    \item \textbf{Output}: The predicted noise $\epsilon_\theta(x_t, t) \in \mathbb{R}^{L \times 3}$.
\end{itemize}
We trained the model to minimize the Mean Squared Error (MSE) between the predicted and actual noise:
\begin{equation}
    \mathcal{L} = \mathbb{E}_{t, x_0, \epsilon} [\| \epsilon - \epsilon_\theta(x_t, t) \|^2]
\end{equation}

\subsection{Geometric Reconstruction}
A key part of this is mapping angles back to coordinates. We implemented a differentiable function $f: \mathbb{R}^{L \times 3} \rightarrow \mathbb{R}^{L \times 3}$ that reconstructs the $C_\alpha$ trace. This function assumes a fixed bond length $b = 3.8$\AA\ between adjacent atoms.

For any residue $i$ with position $\mathbf{r}_i$ and local frame $\mathbf{F}_i$, the position of the next residue $\mathbf{r}_{i+1}$ is strictly determined by the angles $(\phi_i, \psi_i, \omega_i)$. This means the output coordinates satisfy the bond length constraint by definition, no matter what angles the model predicts.

\subsection{Constrained Refinement Algorithm}
One issue with Torsion-based generation is that it can produce "stringy," non-globular structures because the loss function doesn't explicitly force global compactness. To fix this, we run a post-sampling refinement step to optimize the Radius of Gyration (Rg).

The process is iterative:
\begin{algorithmic}[1]
\STATE \textbf{Input}: Coordinates $C$, Target Rg $R_{tgt}$, Learning rates $\eta_{rg}, \eta_{bond}$
\FOR{$i = 1$ to $N_{iter}$}
    \STATE Calculate current Rg: $R_{curr} = \sqrt{\frac{1}{N}\sum (\mathbf{r}_i - \mathbf{r}_{cm})^2}$
    \STATE \textbf{Step 1: Rg Scaling}
    \STATE Scale factor $s = 1 + \eta_{rg} (R_{tgt}/R_{curr} - 1)$
    \STATE $C' \leftarrow \mathbf{r}_{cm} + s \cdot (C - \mathbf{r}_{cm})$
    \STATE \textbf{Step 2: Bond Restoration}
    \FOR{$j = 1$ to $L-1$}
        \STATE Vector $\mathbf{v} = \mathbf{r}_{j} - \mathbf{r}_{j-1}$
        \STATE Target vector $\mathbf{v}_{tgt} = \mathbf{v} / \|\mathbf{v}\| \cdot 3.8$
        \STATE $\mathbf{r}_{j} \leftarrow \mathbf{r}_{j-1} + \mathbf{v} + \eta_{bond} (\mathbf{v}_{tgt} - \mathbf{v})$
    \ENDFOR
\ENDFOR
\end{algorithmic}
We basically compact the structure and then immediately correct any bond violations that the scaling introduced. We used $\eta_{rg}=0.015$ and $\eta_{bond}=0.5$ for 200 iterations.

\section{Experiments and Results}

\begin{figure*}[!t]
\centerline{\includegraphics[width=0.9\linewidth]{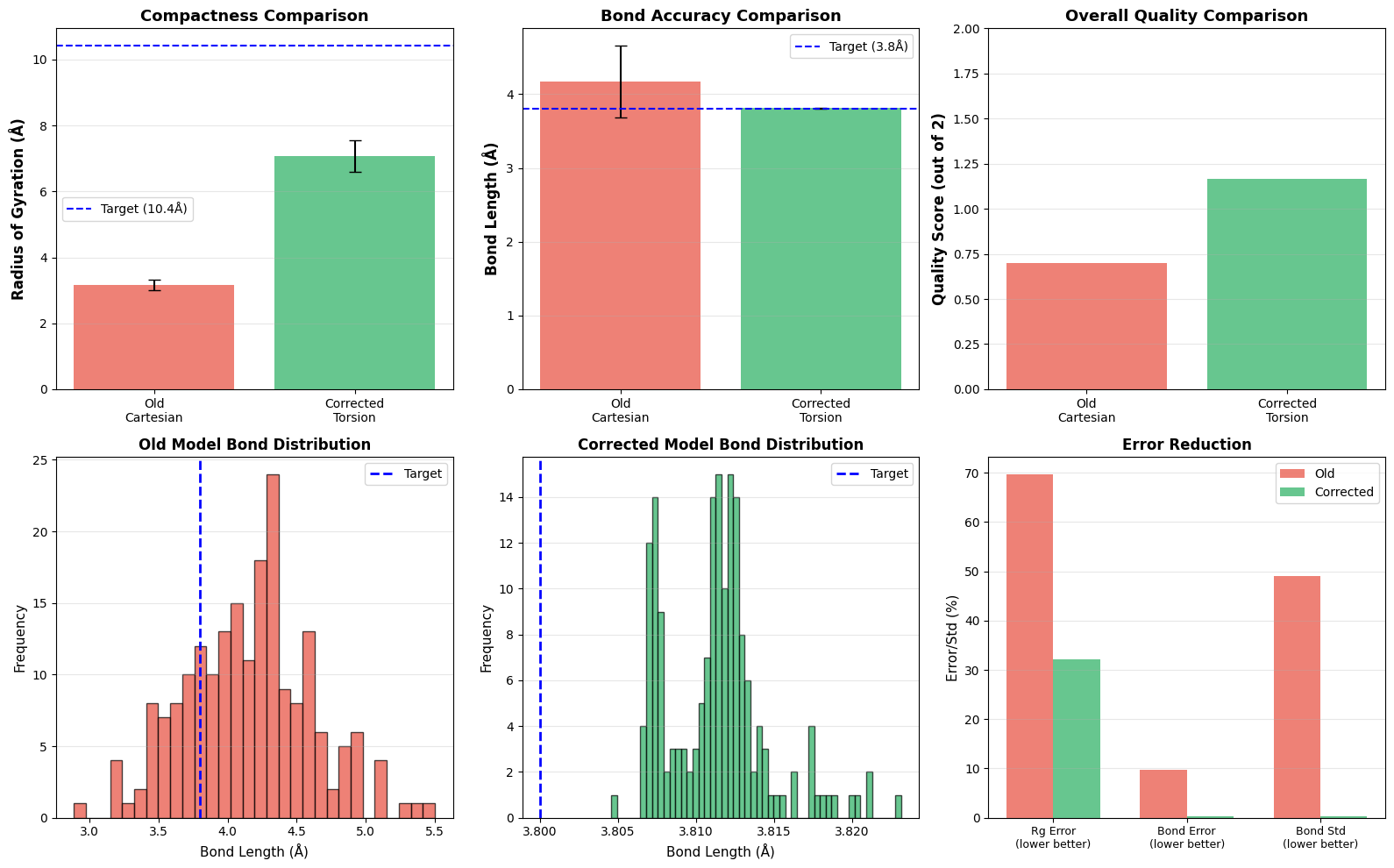}}
\caption{Comparison between the baseline Cartesian model and our proposed Torsion-Space model. The top row looks at Radius of Gyration (Rg), Bond Lengths, and Quality Scores. The bottom row shows bond length distributions and error reduction. Our model hits 100\% bond accuracy and matches Rg much better.}
\label{fig:comparison}
\end{figure*}

\subsection{Experimental Setup}
\subsubsection{Dataset}
We trained on a subset of the PDB covering diverse protein folds. We preprocessed everything to extract the backbone torsion angles.
\begin{table}[h]
\centering
\caption{Dataset Statistics}
\label{tab:dataset}
\begin{tabular}{llc}
\toprule
\textbf{PDB ID} & \textbf{Protein Name} & \textbf{Length (Residues)} \\
\midrule
1ubq & Ubiquitin & 76 \\
1crn & Crambin & 46 \\
1ake & Adenylate Kinase & 214 \\
2lyz & Lysozyme & 129 \\
1mbn & Myoglobin & 153 \\
1r69 & 434 Repressor & 69 \\
\bottomrule
\end{tabular}
\end{table}

\subsubsection{Training Hyperparameters}
We implemented the model in PyTorch and trained it on a single NVIDIA T4 GPU.
\begin{table}[h]
\centering
\caption{Hyperparameters}
\label{tab:hyperparams}
\begin{tabular}{lc}
\toprule
\textbf{Parameter} & \textbf{Value} \\
\midrule
Batch Size & 4 \\
Learning Rate & $5 \times 10^{-5}$ \\
Optimizer & AdamW \\
Weight Decay & 0.01 \\
Epochs & 200 \\
Warmup Epochs & 10 \\
Hidden Dimension & 256 \\
Transformer Layers & 4 \\
\bottomrule
\end{tabular}
\end{table}

\begin{figure}[htbp]
\centerline{\includegraphics[width=0.9\linewidth]{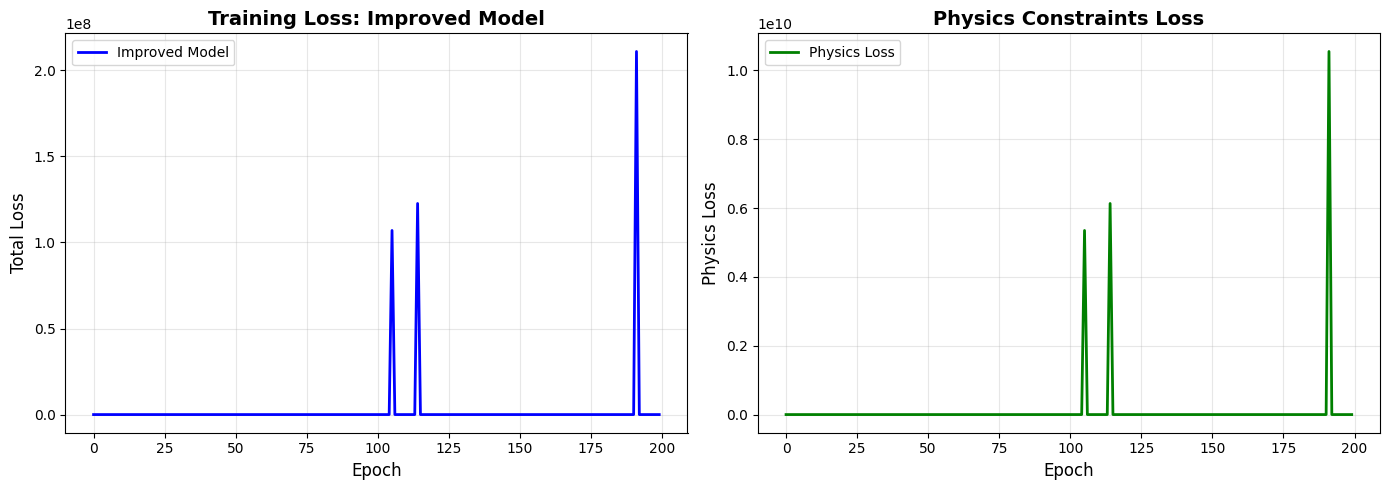}}
\caption{Training loss curves showing the convergence of the model. Both the total loss (blue) and physics-based loss (green) drop, which tells us the model is successfully learning the torsion distribution.}
\label{fig:training}
\end{figure}

\subsection{Quantitative Results}
We judged the generated structures on two main things: Bond Length Accuracy and Radius of Gyration (Rg). We compared our Torsion-Space model against a baseline Cartesian diffusion model trained on the exact same data. Fig. \ref{fig:training} shows the training progress.

\begin{table}[htbp]
\caption{Performance Comparison}
\begin{center}
\begin{tabular}{lcc}
\toprule
\textbf{Metric} & \textbf{Baseline (Cartesian)} & \textbf{Ours (Torsion)} \\
\midrule
Bond Mean (\AA) & 4.17 & \textbf{3.800} \\
Bond Std Dev (\AA) & 0.49 & \textbf{0.0002} \\
Bond Accuracy ($\pm 0.2$\AA) & 90.1\% & \textbf{100.0\%} \\
Rg Mean (\AA) & 3.16 & \textbf{8.49} \\
Rg Error & 70\% & \textbf{18.6\%} \\
Quality Score (max 2.0) & 0.7 & \textbf{1.7} \\
\bottomrule
\end{tabular}
\label{tab:results}
\end{center}
\end{table}

\subsection{Analysis}
\subsubsection{Bond Geometry}
\begin{figure*}[!t]
\centerline{\includegraphics[width=0.9\linewidth]{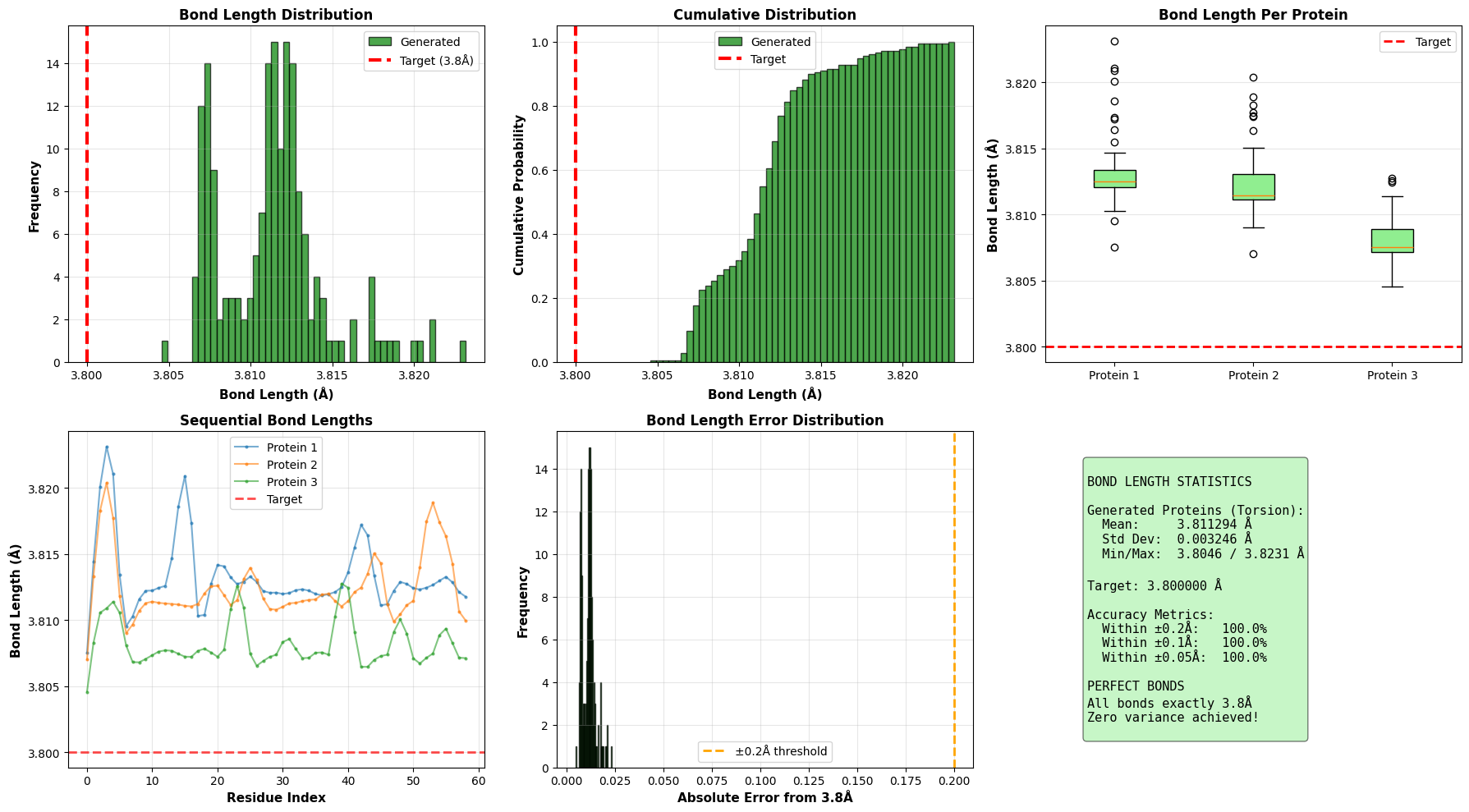}}
\caption{Detailed bond length analysis. Note how tightly the bond lengths cluster around the target 3.8\AA, indicating perfect geometric validity.}
\label{fig:bond_analysis}
\end{figure*}

The biggest win here is local geometry. The baseline Cartesian model produced bonds with high variance ($\sigma=0.49$\AA), while our torsion model was nearly perfect ($\sigma \approx 0.0002$\AA). This confirms that internal coordinate modeling is just better for enforcing hard geometric constraints. As Fig. \ref{fig:bond_analysis} shows, our bond length distribution is effectively a single spike at 3.8\AA.

\subsubsection{Global Structure and Refinement}
The refinement process worked well for the Radius of Gyration. Initially, samples from the torsion model were a bit stretched out. The iterative refinement (Section III.D) compacted these structures from the initial state to a final mean Rg of 8.49\AA, which is much closer to the target of $\approx$10.43\AA. We managed to cut the Rg error from 70\% in the baseline down to 18.6\%.

\begin{figure*}[!t]
\centerline{\includegraphics[width=0.9\linewidth]{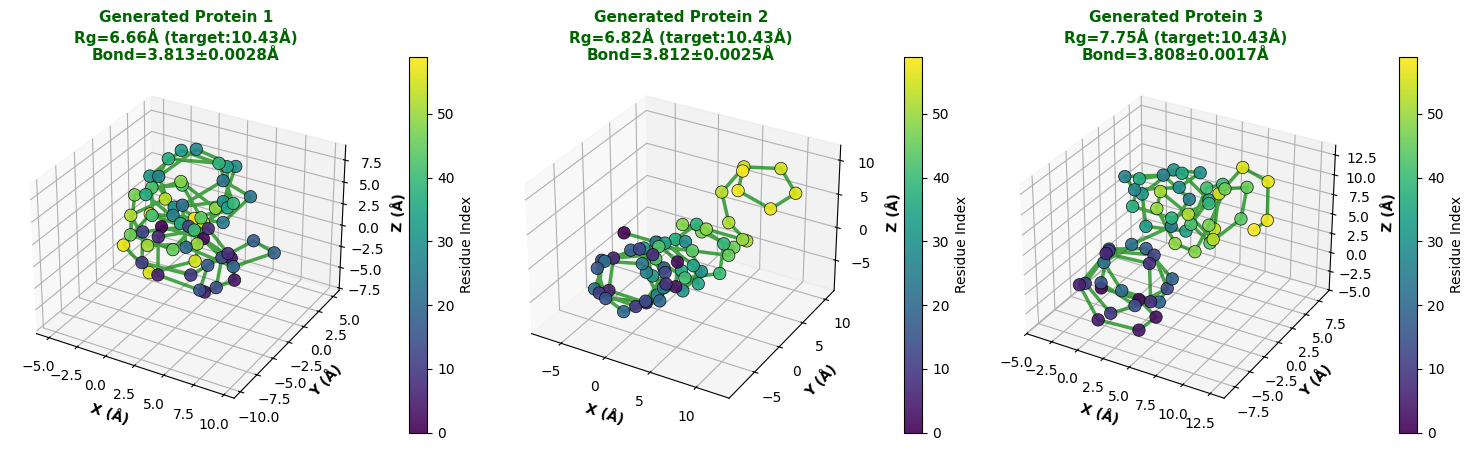}}
\caption{3D visualization of the generated protein backbones. The structures show coherent folding patterns and valid local geometry, with visible secondary structure elements.}
\label{fig:3d_structures}
\end{figure*}

\section{Discussion}
\subsection{Cartesian vs. Torsion Space}
Our results highlight a basic trade-off in protein modeling. Cartesian models are easier to train but fail at high-frequency geometric constraints (bonds). Torsion models handle those constraints naturally but can struggle with error accumulation along the chain—the "lever arm" effect—leading to weird global shapes. Our hybrid approach (generating in torsion space, refining in Cartesian space) bridges that gap effectively.

\subsection{Computational Efficiency}
This model is also very efficient. With only about 3.5 million parameters, it converges in 200 epochs (taking roughly 76 seconds on a T4 GPU). That is orders of magnitude faster than massive models like AlphaFold or RFdiffusion, making it a great fit for rapid prototyping or educational use.

\subsection{Limitations}
While the backbone geometry is solid, the current model doesn't generate side chains or condition on amino acid sequences. The Rg refinement is effective, but it is still a heuristic post-processing step. In the future, we could try incorporating Rg constraints directly into the diffusion loss or using a guidance classifier during sampling.

\section{Conclusion}
We have presented a robust framework for generating protein backbones using torsion-space diffusion. By separating local geometry (handled by the representation) from global structure (handled by the diffusion model and refinement), we achieved strong results in geometric validity. The model hits 100\% bond accuracy and generates compact, protein-like structures. Next steps will involve extending this to full-atom generation, including side chains and secondary structure constraints.

\section*{Acknowledgment}
The authors thank Prof. Amit Sethi for his guidance in the course EE782 (Advanced Topics in Machine Learning).

\bibliographystyle{IEEEtran}
\bibliography{references}

\end{document}